\documentclass[aps,prb,twocolumn,groupedaddress,showpacs]{revtex4-1}
		
		\usepackage{graphicx}
		\usepackage{amsmath}
        \usepackage[usenames,dvipsnames]{color}
        \usepackage{pdfcolmk}

\begin{document}

        \title{Anderson Localization of Electrons in Silicon Donor Chains}

\author{Amintor Dusko}

\affiliation{Instituto de F\'{\i}sica, Universidade Federal do Rio de Janeiro, Caixa Postal 68528, 21941-972 Rio de Janeiro, Brazil}

\author{A. L. Saraiva}

\affiliation{Instituto de F\'{\i}sica, Universidade Federal do Rio de Janeiro, Caixa Postal 68528, 21941-972 Rio de Janeiro, Brazil}

\author{Belita Koiller}

\affiliation{Instituto de F\'{\i}sica, Universidade Federal do Rio de Janeiro, Caixa Postal 68528, 21941-972 Rio de Janeiro, Brazil}

		\date{\today}
		
\begin{abstract}

We construct a model to study the localization properties of nanowires of dopants in silicon (Si) fabricated by precise ionic implantation or STM lithography.
Experiments have shown that Ohm's law holds in some cases, in apparent defiance to the Anderson localization theory in one dimension. We investigate how valley interference affects the traditional theory of electronic structure of disordered systems. Each isolated donor orbital is realistically described by multi-valley effective mass theory (MV-EMT). We extend this model to describe chains of donors as a linear combination of dopant orbitals.
Disorder in donor positioning is taken into account, leading to an intricate disorder distribution of hoppings between nearest neighbor donor sites (donor-donor tunnel coupling) -- an effect of valley interference. The localization length is obtained for phosphorous (P) donor chains from a transfer matrix approach and is further compared with the chain length.
We quantitatively determine the impact of uncertainties $\delta R$ in the implantation position relative to a target and also compare our results with those obtained without valley interference. We analyse systematically the aimed inter-donor separation dependence ($R_0$) and show that fairly diluted donor chains ($R_0=7.7$ nm) may be as long as 100 nm before the effective onset of Anderson localization, as long as the positioning error is under a lattice parameter ($\delta R <0.543$ nm).

\end{abstract}

		\pacs{03.67.Lx, 85.30.-z, 85.35.Gv, 71.55.Cn, 73.20.-r}

		\maketitle
		
		\section{\label{sec:Introduction}Introduction}

On demand fabrication of dopant arrangements in silicon is made possible by recent STM lithography techniques. Nanowires constructed with this technique obey Ohm's law, in apparent defiance to the Anderson localization theorem.~\cite{Weber2012,Weber2014} Even in quasi-one dimensional systems constructed through ionic implantation of donors, for which positional disorder is significantly larger, metal-insulator transition is observed at large enough dopant densities.~\cite{Shinada2005,Prati2012} It is unclear what aspects of the host semiconductor drive this behavior and what can we do to effectively engineer these transport properties.

We consider here a system of dopants deliberately implanted at target positions so as to regulate their tunnel coupling (or hopping). Our model incorporates the currently unavoidable imprecisions in donor positions and, in particular, how they manifest in the hopping distribution. We develop a theory of linear combinations of dopant orbitals (LCDO), in which each donor orbital is described within multi-valley effective mass theory (MV-EMT) including a central cell correction. We then obtain the transfer matrix for donor chains with realistic positioning disorder models and extract the localization length for finite chains based on the asymptotic behavior of the Lyapunov exponent of the transformation defined by it. We compare systems with and without valley interference, showing that valleys play two opposite roles in the localization of states -- it increases the number of states available at the point of charge neutrality(half-filled band) and introduces the possibility of fully destructive interference, leading to a broken link in the chain and reducing the paths through which the current can percolate.
We also study effects of uncertainty in positioning in two and three dimensions (a disk or a sphere of uncertainty), consistent with bottom-up STM lithography and top-down ionic donor implantation methods, respectively.

\section{\label{sec:Formalism} Donors in silicon within effective mass}

We construct here the model for the electronic structure of dopant chains extending the EMT of donor impurities in Si. First, we revise the central cell corrected model of an impurity in Silicon within effective mass (Sec.~\ref{sec:KL}). We also revise how to construct the wavefunction of a donor pair from a single dopant orbital, in analogy with the molecular orbital theory. Then this analogy is extended into the model of a chain of atoms, with a wavefunction described as a linear combination of central cell corrected Kohn-Luttinger wavefunctions centered at the impurity sites, which we dub LCDO.

\subsection{\label{sec:KL} Single donor}

A distinct feature of the Si band structure is the six-fold degeneracy of the conduction band edge, located at ${\bf k}_\mu$ along the $\langle 100 \rangle$  directions, $\mu =\pm x,\pm y, \pm z$ with $|{\bf k}_\mu|=k_0=0.85\left({\frac{2\pi}{a_{\rm Si}}}\right)$ in the FCC Brillouin zone ($a_{\rm Si}=0.5431$ nm is the conventional FCC lattice parameter for Si).~\cite{Madelung2012} For bulk Si, the periodic hamiltonian does not couple the valleys, which remain degenerate.
A substitutional donor breaks the translational symmetry and the steep donor potential couples different valleys, resulting in a non-degenerate ground state to which the six valleys contribute equally (A$_1$ symmetry). The A1 state is well separated (12 meV) from the 1st excited state.

Kohn and Luttinger\cite{Luttinger1955} (KL) proposed, within effective mass theory (EMT), a ground state variational wave-function pinned at donor position with the correct A$_1$-symmetry written in terms of hydrogenic envelopes and Bloch functions for each conduction band minimum. For a donor at ${\bf {r}}=0$,
\begin{equation}
\Psi_{\rm KL} ({\bf r}) ={\frac{1}{\sqrt{6}}} \sum_{\mu=1}^6 F_\mu ({\bf r}) e^{i {\bf k}_\mu \cdot {\bf r}} u_\mu (\bf r)
\label{eq:wave-function}
\end{equation}
where $F_\mu (\bf r)$ are hydrogenic-like envelope functions and $u_\mu (\bf r)$ are the periodic parts of the six Bloch functions. The effective mass anisotropy affects the ground state envelope functions, suggesting the use of deformed 1s orbitals with $a$ and $b$ as variational parameters,
\begin{equation}
\label{eq:envelopes}
F_{\pm z} ({\bf r}) ={\frac{1} {\sqrt{\pi a^2 b}}} \, \exp\left({-\sqrt{{\frac{x^2+y^2} {a^2}}+{\frac{z^2} {b^2}}}}\right),
\end{equation}
and equivalently to $F_{\pm x}$ and $F_{\pm y}$. A similar variational envelope was proposed by Kittel and Mitchell.~\cite{Kittel1954}

This gives a species-independent description and a six-fold degenerate donor ground state -- in contrast, experiments show a variation in binding energies among different group V dopants (P, As, Sb, Bi) and a non-degenerate ground state as mentioned before. Both problems are corrected by a species-dependent central cell potential, which accounts for the more attractive potential close to the donor site, where the screening by the core electrons is less effective. Besides splitting  the 1S manifold into a non-degenerate A1 state and 3- and 2-fold degenerate excited states (T2 and E, respectively) at the experimentally observed energies, the  main consequence of this correction to the present study is to contract the donor ground state wavefunction -- an effect confirmed experimentally.~\cite{saraiva2016} The central cell prescription discussed in Ref.~\onlinecite{saraiva2015} is incorporated in our model calculations below.

\subsection{\label{array} Donor pairs and linear arrays}

In analogy with the standard linear combination of atomic orbitals (LCAO) scheme in quantum chemistry, we construct the wavefunction of a pair of donors as a linear combination of dopant orbitals (LCDO) described in Eq.~(\ref{eq:wave-function}) centered at the substitutional donor sites $\mathbf{r_1}$ and $\mathbf{r_2}$. The dopant pair (molecule) variational wavefunction reads
\begin{equation}
\Psi_{\rm mol}({\bf r})=\alpha_1 \Psi_{\rm KL} ({\bf r}-{\bf r_1})+\alpha_2 \Psi_{\rm KL} ({\bf r}-{\bf r_2})
\label{eq:wave-function}
\end{equation}
and the coefficients $\alpha_1$ and $\alpha_2$ must be determined variationally, under the normalization constraint $\langle\Psi_{\rm mol}|\Psi_{\rm mol}\rangle=1$. As in the LCAO procedure, this leads to a set of Rothaan equations for the coefficients,~\cite{slater1963book} which can be written as a Fock matrix ${\mathbf F}$ and an overlap matrix ${\mathbf S}$. Here we are interested in the single particle effects, so that the Fock operator is  the single electron Hamiltonian.

This approach is valid as long as \emph{i)} the interdonor distance $\mathbf{R_{12}}=\mathbf{r_2}-\mathbf{r_1}$ is not too small, so that the assumptions of EMT are still valid (continuum approximation); and \emph{ii)} the ground state wavefunction is still mainly composed of a symmetric combination of valley states pinned around each site. The latter assumption is more restrictive -- while the actual ground state combination of valleys slowly changes as the donors are brought closer together,~\cite{klymenko2014} near a distance $R_{12}\approx6$~nm there is a sudden change from mostly A$_1$-like to mostly T$_2$-like combination of valleys.~\cite{klymenko2014} Therefore, our model is qualitatively inaccurate below this interdonor distance.

If we call $|1\rangle$ and $|2\rangle$ the orbitals centered in each donor, the pair Hamiltonian within a one-electron LCDO description may be written as
 \begin{equation}
        \label{eq:pair}
		\widehat{H}_{\rm pair}= \epsilon\hspace{1pt} \big[ |1\rangle \langle 1| + |2\rangle \langle 2|\big] + t \big[ |1\rangle\langle 2| +|2\rangle\langle 1|\big],
		\end{equation}
where $\epsilon$ is the isolated donor eigenenergy, which may be taken as the energy origin and $t$ is the tunneling or hopping energy which is a function of the donors relative position ${\bf R}_{12}$. These LCDO parameters are explicitly obtained semi-empirically -- as discussed below. This procedure may be extended for larger sets of dopants.

We formulate a simple one-orbital per site LCDO Hamiltonian for linear arrays of well separated donors based on nearest-neighbor pairs  considerations. To describe the array's ground state properties it remains acceptable to restrict the basis set to the ground state orbital in each donor, since the perturbation due to nearby donors is relatively small. Further approximations aiming at simplifying the numerical LCDO parameters calculations are discussed in Sec.\, \ref{sec:Approximations}.

Following (\ref{eq:pair}), we write a nearest-neighbors LCDO Hamiltonian in the basis set of the A1 KL variational solutions, as given in Eq.(\ref{eq:wave-function}), to describe the ground state of a linear array of  hidrogenic-like impurities,
	    \begin{equation}
        \label{eq:Hamiltonian}
		\widehat{H}=\sum_{\langle i,j\rangle} \left[ \frac{1}{2} \epsilon \hspace{1pt} n_{i} +  t_{i,j}\left(c_i^{\dagger} c_j + c_j^{\dagger} c_i \right) \right],
		\end{equation}
where $\epsilon$ is the on-site energy (site-independent for a single donor species) and $t_{i,j}$ is the nearest neighbor hopping,  $t$ in Eq.\,(\ref{eq:pair}), for donors at sites $i$ and $j$. We allow the tunnel coupling to be dependent on the particular pair, incorporating possible displacements of donors relative positions along the fabrication procedure, so that $t_{i,j}=t({\bf R}_{ij})$.
The creation ($c^{\dagger}_i$) and annihilation ($c_i$) operators refer to the occupation of the  orbital $|i\rangle$ at site $i$, thus $n_{i}=c_i^{\dagger} c_i$ is the site occupation operator.

\section{\label{sec:Approximations} Calculation of LCDO parameters - Single \textit{vs} Multi-Valley Model}

It is possible to obtain, within plausible approximations, analytic expressions for the LCDO parameters. The first approximation is to assume isotropic envelopes by taking $a=b=a_{cc}$, where the subscript $cc$ refers to the effective Bohr radius obtained from a central cell corrected potential,\cite{saraiva2015} given in table \ref{tab:parameters}. We do this for each of the valleys separately, preserving the conduction band six-fold degeneracy and the physical insight on valley degeneracy effects, e.g., valley interference. Furthermore, we focus here on dilute doping, which is easily accessible within effective mass (as opposed to the dense limit \cite{Weber2014}). Within this dilute limit, the orbital overlap is small enough to be treated perturbatively. Therefore we dismiss hoppings among donors that are not nearest neighbors.

\begin{table}[ht!]
\begin{center}
\begin{tabular}{|c||c|c|} \hline
Donor & $E_0$ (meV) & $a_{cc}$ (nm)  \\ \hline \hline
P    & 45.58     & 1.106 \\ \hline
As    & 53.77     & 0.815 \\ \hline
Sb    & 42.71     & 1.241 \\ \hline
Bi    & 70.88     & 0.580 \\ \hline
\end{tabular}
\caption{\label{tab:parameters} Single donor ground state binding energy ($E_0$) and effective Bohr radius [See text] for group V dopants as given and described in Ref.\hspace{2pt}\onlinecite{saraiva2015}.}
\end{center}
\end{table}

To calculate the matrix elements of the atomistic hamiltonian for a one-dimensional array of $N$ donors $\langle j |\hat{H}|i\rangle$, we rewrite $\hat{H}= \hat{H}_i+\hat{H}'$, where  $\hat{H}'=\sum_{j\neq i}\hat{V}_j$ is the perturbation potential due to all other cores and $\hat{H}_i=\hat{K}+\hat{V}_i$ consists of the kinetic energy ($\hat{K}$) plus the core potential $\hat{V}_i$ for donor $i$. This is useful because, by construction, $\hat{H}_i|i\rangle = -E_0 \,|i\rangle$ where $E_0$ is the single donor binding energy.

Since $\langle j|\hat{H}'|i\rangle \ll  \langle j |\hat{H_i}|i\rangle$, the perturbation term is neglected in the hopping energy expression, resulting
 \begin{subequations}
\label{eq:LCDOparameters}
\begin{eqnarray}
 t_{ij}=& \langle j |\hat{H_i}|i\rangle +\langle j|\hat{H}'|i\rangle, \\  \approx & - E_0\hspace{1pt}S(R_{ij},a_{cc}).
\end{eqnarray}
\end{subequations}
So $t_{ij}$ is a function of the single donor ground state  binding energy ($E_0$) and the overlap between $|i\rangle$ and $|j\rangle$ nearest neighbors orbitals $S(R_{ij},a_{cc})=\langle i | j \rangle$.

If energies are measured from the bottom of the conduction band, we would identify $\epsilon$ with  $-E_0$. Instead, we set the origin of energies to the center of the impurity band setting $\epsilon = 0$, without loss of generality.

In order to highlight valley interference effects in our one-electron-one-orbital isotropic envelopes model, we consider two expressions for the hopping (we omit the $i,j$ labels when no ambiguity is raised).
   Firstly we take an expression neglecting valley interference efects, equivalent to the dopant orbitals appearing on single valley semiconductors,\cite{saraiva2015}
\begin{equation}
	\label{eq:tsv}
	t^{sv}=E_0 \hspace{1pt} e^{-\frac{R}{a_{cc}}}\left(1+\frac{R}{a_{cc}}+\frac{R^2}{3a^2_{cc}}\right).
\end{equation}
In this case $t$ is a function of $|{\bf R}| = R$ only. The expression for the overlap is simply that of an H$_2$ molecule with orbital radius and energies rescaled by the material effective parameters.

\begin{figure}[h]
		\includegraphics[clip,width=\columnwidth]{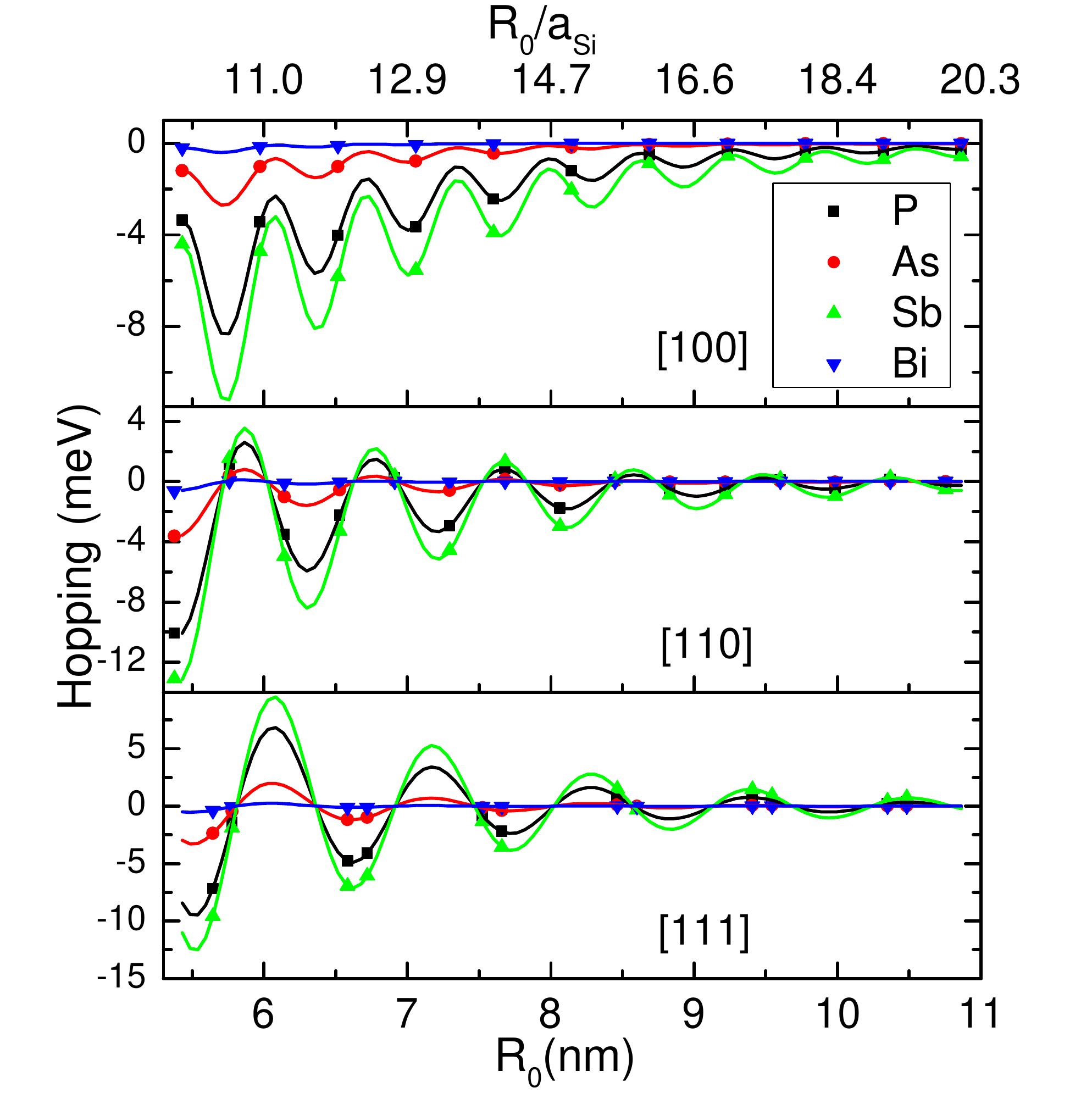}
		\caption{\label{fig:hopping} (Color online) Hopping ($t^{mv}$) as a function of the distance between Group-V (P, As, Sb and Bi) donor pairs along the given crystal directions. Symbols represent the allowed positions in the Si lattice for substitutional donors.
  }
\end{figure}

The second approach considers the six-fold degenerate Si valleys giving, for isotropic envelopes,
    \begin{equation}
	\label{eq:tmv}
	t^{mv}=t^{sv} \left[\frac{1}{3}\sum_{\eta=1}^{3}\cos \left( k_0\hspace{1pt}R_\eta \right)\right],
	\end{equation}
referred to as multi-valley ($mv$) model.  Here $R_\eta$ ($\eta=x,~y,~z$) are the cartesian coordinates of ${\bf R}$ along the Si cubic axes. Interference among the six Bloch functions manifests itself as the term in square brackets, resulting in an oscillatory behavior of the tunnel coupling. Depending on the vector ${\bf R}$, strong suppression of the tunnel coupling may occur. Figure \ref{fig:hopping} presents the hopping energy connecting a pair of donors a distance R$_0$ apart along three crystal directions ([100], [110] and [111]). Note that the more localized Bi donor orbitals leads to a negligible coupling as compared to the other donor species.

For each eigenstate $\Psi(x)$, we define a localization length ($\xi$) from the asymptotic behavior of the wavefunction,
\begin{equation}
\label{eq:loc}
\Psi(x) \rightarrow \exp \left[{-{|x|}/{\xi}}\right].
\end{equation}
This quantity describes the spatial extent of the wavefunction, and thus is closely linked with transport properties. We formally determine this localization length next by the standard transfer matrix approach.

\section{\label{sec:TMApproach}Transfer Matrix Approach}

The Transfer Matrix Approach(TMA) is an efficient way to calculate the localization lengths in 1D chains and is easily extended to quasi-1D structures.~\cite{MacKinnon1981,Pichard1981,Pichard1981B,MacKinnon1983,Kramer1993,Itoh2004}
A very concise description of the operational steps  involved is given here.

We write the eigenfunctions in the basis of A1 orbitals as $|\Psi\rangle = \sum_n \phi_n |n\rangle$, where $n$ is the index for the chain site. The TMA approach\cite{Pichard1981,Evangelou1986,Goldsheid1989} follows from the relation among the wave function amplitudes $\phi_n$, $	t_{n,n-1} \phi_{n-1} + t_{n,n+1} \phi_{n+1} = E \phi_n,$
which is directly cast into a tensorial formulation,
\begin{subequations}
\label{eq:TMequation}
\begin{eqnarray}
\widehat{\Phi}_n=\left(\begin{array}{c}
\phi_{n+1} \\
\phi_{n} \\
\end{array}\right), \qquad \widehat{\Phi}_n=\widehat{T}_n \cdot \widehat{\Phi}_{n-1}, \label{eq:TMequationRel}\\
 \widehat{T}_n =
 \left(\begin{array}{cc}
{ E }/{t_{n,n+1}} & -{t_{n,n-1}}/{t_{n,n+1}}\\
1 & 0\\
\end{array}\right)\label{eq:TMequationDef},\\
\widehat{\Phi}_{L-1}=\left(\prod_{i=1}^{L}\widehat{T}_{L-i}\right) \cdot \widehat{\Phi}_{0}. \label{eq:Mapping}
\end{eqnarray}
\end{subequations}
These relations show that indeed $\widehat T$  acts as a transfer operator for the Lyapunov vectors ($\widehat{\Phi}_n$), while (\ref{eq:Mapping}) is guaranteed to asymptotically converge by the Oseledec theorem.~\cite{Oseledec1968,Ruelle1982,Krengel1985,Carmona2012}

The mapping described in Eq.~(\ref{eq:Mapping}) defines how the wavefunction decays from a reference site $i=0$ into $i=L$. Therefore, we can relate the localization length $\xi$ and the Lyapunov characteristic exponents (LCEs) $\gamma_p$, with $p=0,1$, for both components of the vector. The LCEs are calculated numerically, as discussed in Appendix~\ref{app:LyapExp}.

Once the mapping has converged, any iteration like (\ref{eq:TMequationRel}) gives  $\widehat{\Phi}_{n}^{(p)} = \exp[\gamma_p \cdot \left|{\bf R}_{n}\right|]\cdot\widehat{\Phi}_{n-1}^{(p)}$, where the index $p$ refers to each LCE and $\left|{\bf R}_{n}\right|$ is the distance between consecutive sites $\left(n-1,n\right)$. The LCEs lead to solutions interpreted as either an increasing wavefunction (which we call $\gamma_1$) or decreasing ($\gamma_0$).  The unphysical exponent $\gamma_1$ is discarded, and the decreasing solution leads to a localization length $\xi=|\gamma_0^{-1}|$.

\section{\label{sec:DisorderAndDensity} Donor arrangement and disorder}

We simulate a (quasi-) linear array of substitutional donors in Si along a [110] symmetry direction as sketched in Fig.~{\ref{fig:cartoon}. Target positions are assigned at evenly spaced substitutional atomic sites in the Si structure, with nearest neighbors separated by a vector ${\bf R_0}$. In the absence of disorder, the electronic structure is trivially calculated, since in Eq.(\ref{eq:Hamiltonian}) $t_{i,j}=t({\bf R_0})$, constant for all nearest-neighbors pairs.

However, in real samples, unavoidable deviations from the target positions lead to a disordered Hamiltonian. Here the most affected terms are the off-diagonal matrix elements, $t_{i,j} = t({\bf R_{ij}})$, which specify our model of disorder, as the hopping distribution is implicitly defined by the donors positions.  Positioning disorder is simulated here by a geometric parameter $\delta$ (see Fig {\ref{fig:cartoon}). We assign around each target position $n{\bf R_0}$, with $n$ integer, a region of uncertainty: a disk or a sphere (we discuss both) of radius $\delta$. The donor is randomly positioned at a substitutional site within the uncertainty  region. This approach has the same ingredients found in fabrication methods based on impurity implantation~\cite{Shinada2005,Prati2012,Weber2012} followed by an annealing procedure (guaranteeing the occupation of the energetically favorable substitutional site).

The amount of disorder is dictated by the accuracy of the donor positioning method. Ionic implantation is characterized by a straggle region in the longitudinal direction and some lateral uncertainty due to imperfect collimation of the accelerated ionic beam ~\cite{Shinada2005,Prati2012} -- which typically leads to a three dimensional uncertainty region. If instead the nanostructure is fabricated by bottom-up litographic implantation,~\cite{Weber2012} the uncertainty region is reduced by the far better precision of the litographic intrument tip. Furthermore, the uncertainty is initially confined to the exposed surface - which we model as a two-dimensional uncertainty disk (in fact, often the overgrowth of Si on top of the donor arrangement and thermal treatment lead to some three dimensional uncertainty due to diffusion).
In both cases, the radius $\delta$ establishes the disorder distribution in the electronic hamiltonian. A more general model would allow an ellipsoidal region, but an estimate of the aspect ratio for each experiment would be needed. Instead, we restrict the study to the special cases of a sphere and a disk in order to obtain general features due to the dimensionality of the diffusion, instead of modelling the peculiarities of each method. Only  discrete values of $\delta$, representing a change in the number of lattice sites contained within the sphere/disk are meaningful. For example, all $\delta <a_{\rm Si} \sqrt{3}/4$ are equivalent to $\delta = 0$, since all of them only contain the central target site.

We consider the target chains along the [110] direction, i.e. $(R_0)_x=(R_0)_y=n\,a_{\rm Si}/2$, as in reference [\onlinecite{Weber2012}].  The role of donors' density ($\rho = 1/R_0$) is assessed by considering target  chains with different linear densities, i.e., changing $R_0$.
%
\begin{figure*}
        \includegraphics[clip,width=\textwidth]{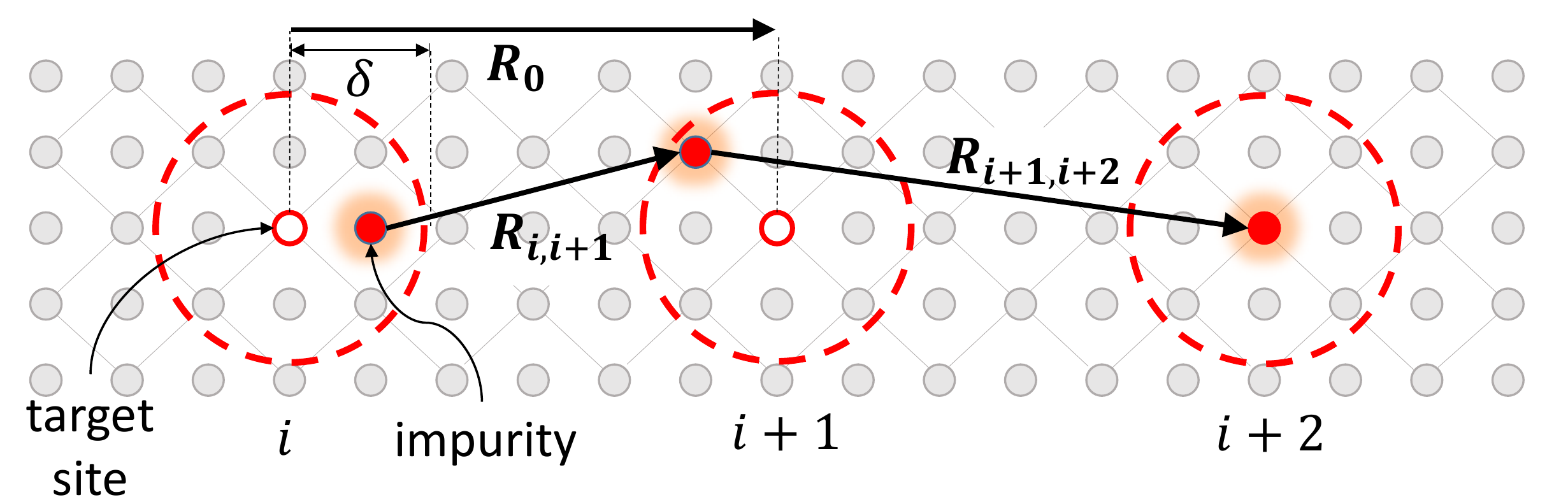}
		\caption{\label{fig:cartoon} (Color online) Schematic representation of our model of disorder: target sites (small circles in solid lines) are aligned along a single [110] axis and separated by ${\bf R_0}$. Dashed circles of radius $\delta$ centered at each target site define the uncertainty region. In 2D the actual donor position is randomly picked among the substitutional sites inside this disk. In the 3D disorder model the position is picked inside a sphere of radius $\delta$. Relative position vectors for consecutive sites are indicated.
  }
		\end{figure*}

 %
\section{\label{sec:Conduction} Mobility edges}

Electronic transport in a chain of donors differs from a regular condensed matter system in many ways. A relevant aspect is the finiteness of these chains, which may have a length comparable or lesser than the extent of an exponentially localized state. In the former case, such electronic state would contribute to charge currents and, in the absence of scattering processes, it conducts in the so-called ballistic transport regime. In the presence of relaxation phenomena such as electron-phonon scattering, transport would fit into the Drude or ohmic transport picture. Our criterion for metallic conductivity is based on these considerations, and we do not explicitly include relaxation.

The mature theory of electronic localization by disorder is discussed extensively in the literature. For infinite 3D disordered systems the theoretical prediction is that, while  extended and localized states co-exist in the solid, they are completely segregated in energy.  Eigenenergies of localized (toward the band edges) and of  delocalized (around the band center) states are sharply separated at the so-called mobility edges $(\mu)$.~\cite{Anderson1958,Bulka1985}

In 1D, disorder of any strength localizes all states. Considering that metallic-like current involves  transport by delocalized electrons, it should not be observed in disordered chains, where no mobility edge is present. However, when dealing with {\em finite} arrays, we may define an effective mobility edge, $\mu_{e\!f\!f}(L) $, as the value of the energy at which the localization length of the states becomes larger than the chain length $L$, a scenario compatible with current flow along a disordered 1D system.

In 1D, our nearest-neighbors model with off-diagonal disorder gives a symmetric distribution of eigenstates with respect to the on-site energy, chosen here as $\epsilon = 0$, so that for $\xi < L$ in all energy domain we get $\mu_{eff} = 0$.

It is also convenient from the theoretical point of view to define the converse quantity. We call length edge the maximum length $\Lambda$ that may be ascribed for a chain characterized by the interdonor spacing $R_0$ and disorder $\delta$ before transport no longer occurs.

In the results presented here, each data point is obtained from averages over an ensemble of $N$ chains of equal length $L$. In all chains, donors are positioned according to the geometrical parameters $\delta$ and $R_0$. The ensembles were large enough to get convergence in the mean value of the localization length better than 1\%, which required $N  \approx 10^4$.

  Figure \ref{fig:Densities} shows the length edge dependence on the interdonor separation ($\Lambda$ vs $R_0$) for a fixed disorder $\delta$ = 0.4 nm. It is possible to observe a non-monotonic fluctuation  behavior, with more pronounced  and steeper oscillations for 2D than for 3D disorder. This counter-intuitive result may be understood analyzing the impact of the dimensionality on the number of possible atomic arrangements of the chain.

   In the 2D disorder model, this radius $\delta$ contains 5 possible sites for an impurity -- thus 25 relative positions for consecutive donors -- among which only 9 are inequivalent (lead to different tunnel couplings).
     In the 3D case with the same value of $\delta$, there are 17 sites, leading to 289 different relative positions of consecutive pairs, distributed in 36 inequivalent distances.

More generally, for a fixed disorder radius $\delta$, there are more possibilities for different vectors $R_ij$ in 3D than in 2D and, as a result, fluctuations in $\Lambda$ tend to average out becoming less pronounced in the 3D case, as illustrated in Fig. \ref{fig:Densities}. Reduced  fluctuations in 3D with respect to otherwise equivalent models in 2D  are also observed in Figs.~(\ref{fig:zeta} and~\ref{fig:AllDos-77A}), as discussed next.

		\begin{figure}[h!]
		\includegraphics[width=\columnwidth]{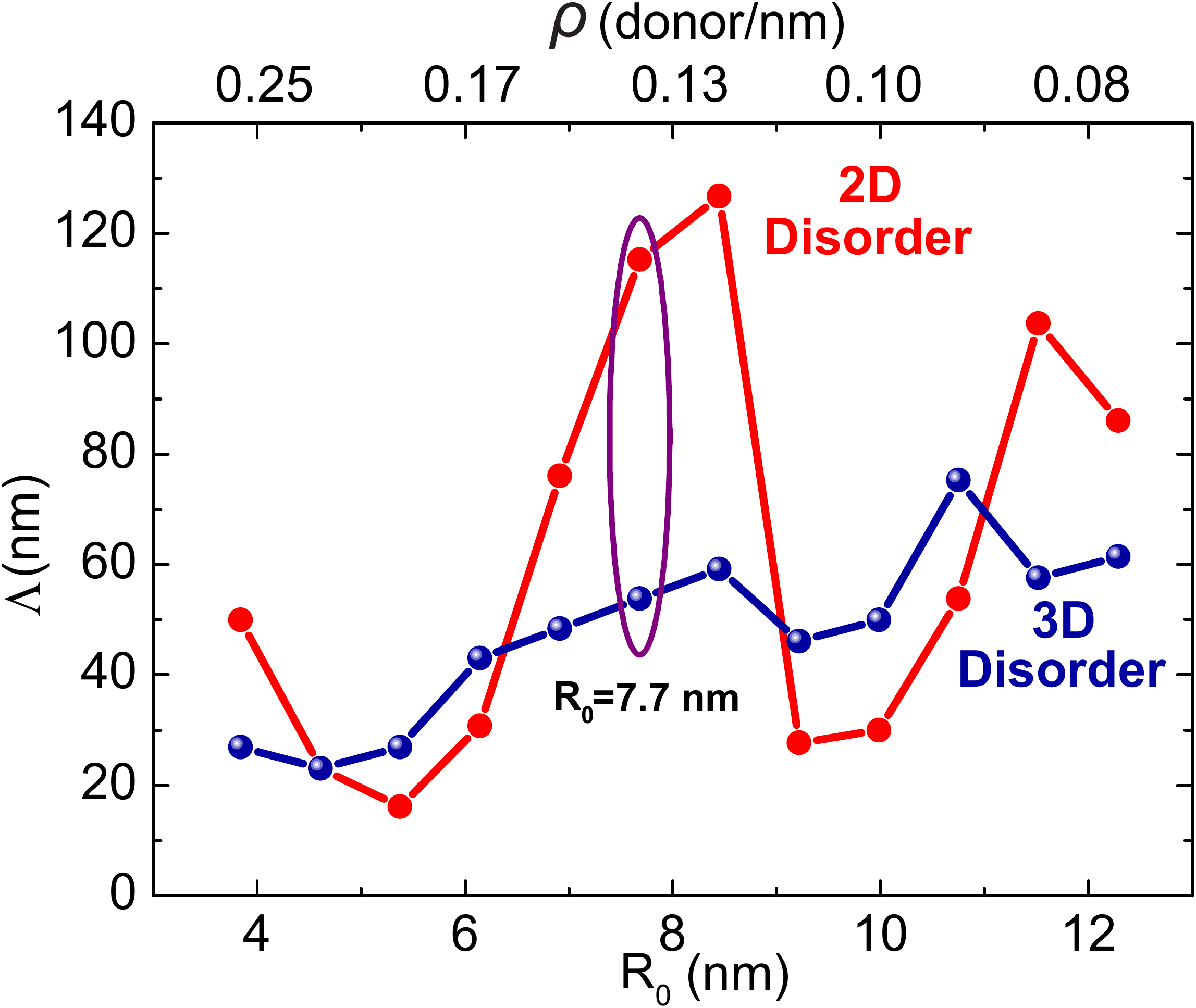}
		\caption{\label{fig:Densities} Length edge ($\Lambda$)--the upper limit of length for a chain to sustain electronic transport--as a function of distance between consecutive target positions ($R_0$) in 2D and 3D disorder models for  multi-valley hopping and $\delta=0.4$ nm. The encircled region is expanded in Fig. 2 which shows the dependence of $\Lambda$ on disorder $\delta$ for $R_0=7.7$ nm. The region below each line corresponds to the current-carrying behavior. Error bars are smaller than the data points
  }
		\end{figure}

The effect of increasing disorder on the maximum chain length $\Lambda$ is investigated in Fig.~\ref{fig:zeta} for fairly dilute donor chains ($R_0=7.7$ nm). Here we note a decrease in $\Lambda$ with increasing disorder, a plausible result. Again, the 2D data points show distinct fluctuations as compared to the smoother 3D cases.

 We discuss general trends in $\Lambda$ in sec. \ref{sec:Trends}.
		\begin{figure}[h!]
		\includegraphics[width=\columnwidth]{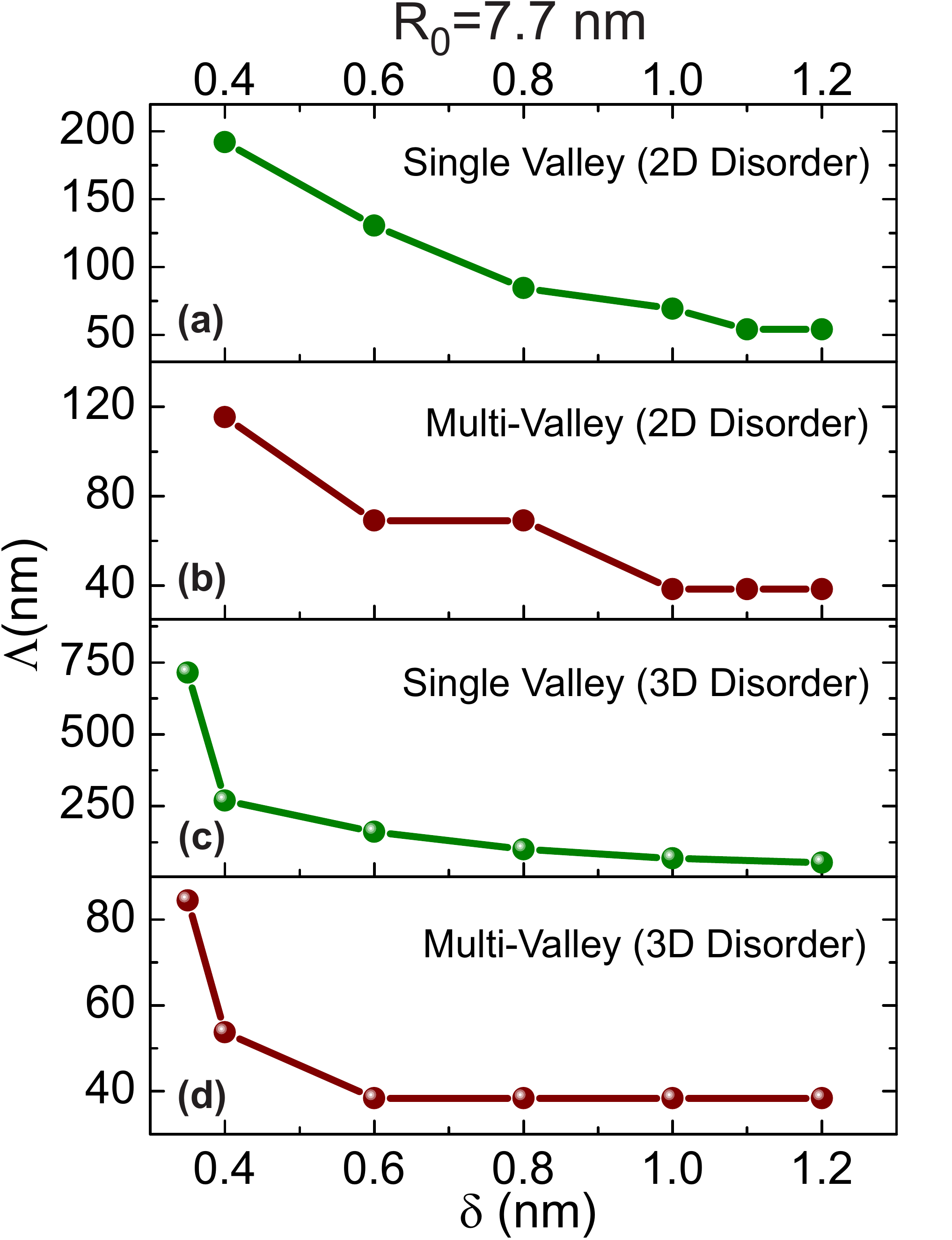}
		\caption{\label{fig:zeta}(Color online)
Disordered chain length edge ($\Lambda$) as a function of 2D(3D) disorder radius ($\delta$) for single valley model [(a) and (c)] and multi-valley model [(b) and (d)]. Results are presented for a distance  $R_0=7.7$ nm between consecutive target implantation points.
  }
		\end{figure}

		\begin{figure}[h!]
		\includegraphics[width=\columnwidth]{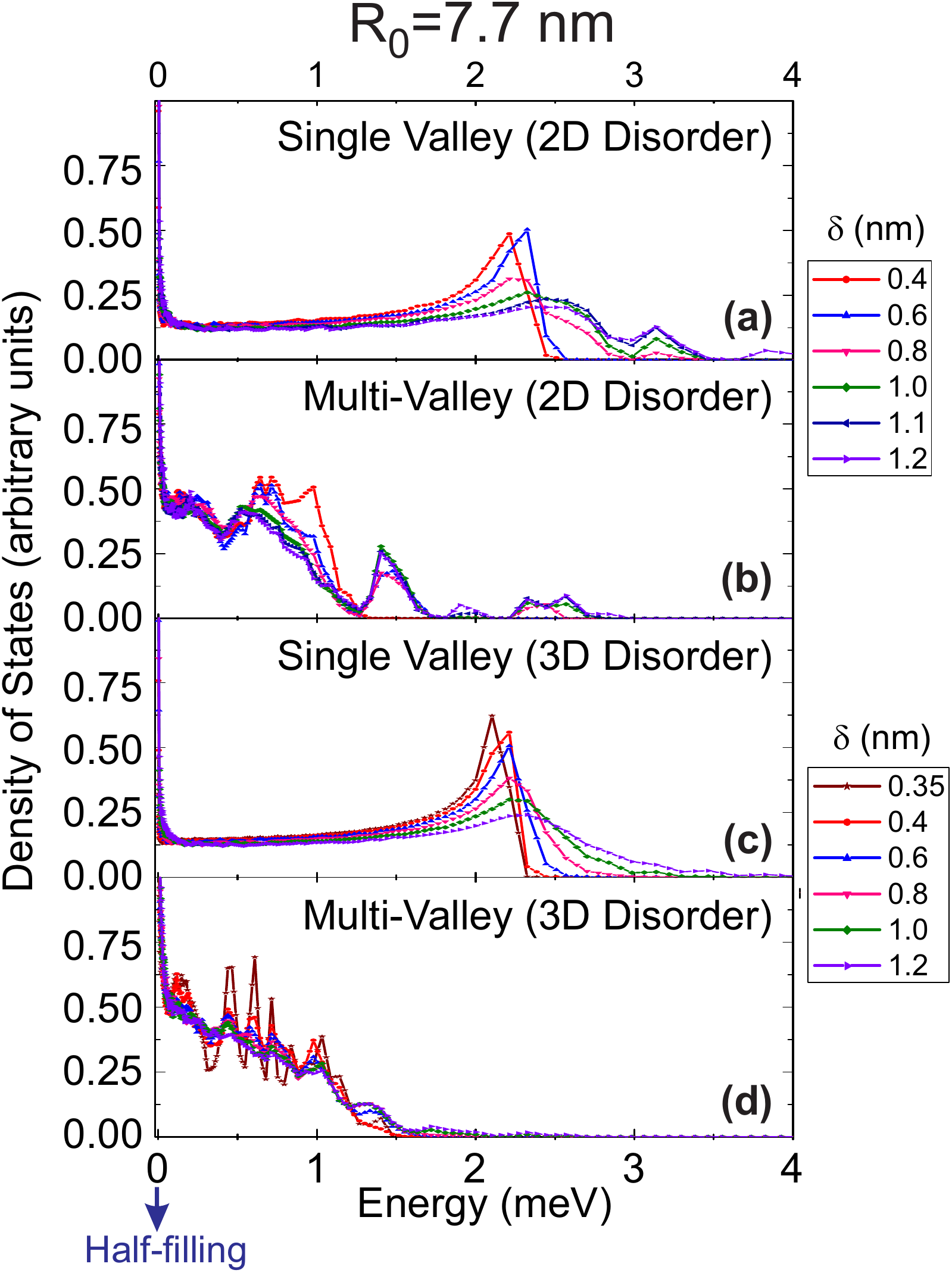}
		\caption{\label{fig:AllDos-77A}(Color online)
Density of states in energy for six different disorder radii ($\delta$). Our model leads to a single band which is symmetric around the on-site energy $\epsilon$, taken to be zero here. The model (SV or MV, 2D or 3D) is indicated in each frame.
  }
		\end{figure}

		\begin{figure}[h!]
		\includegraphics[width=\columnwidth]{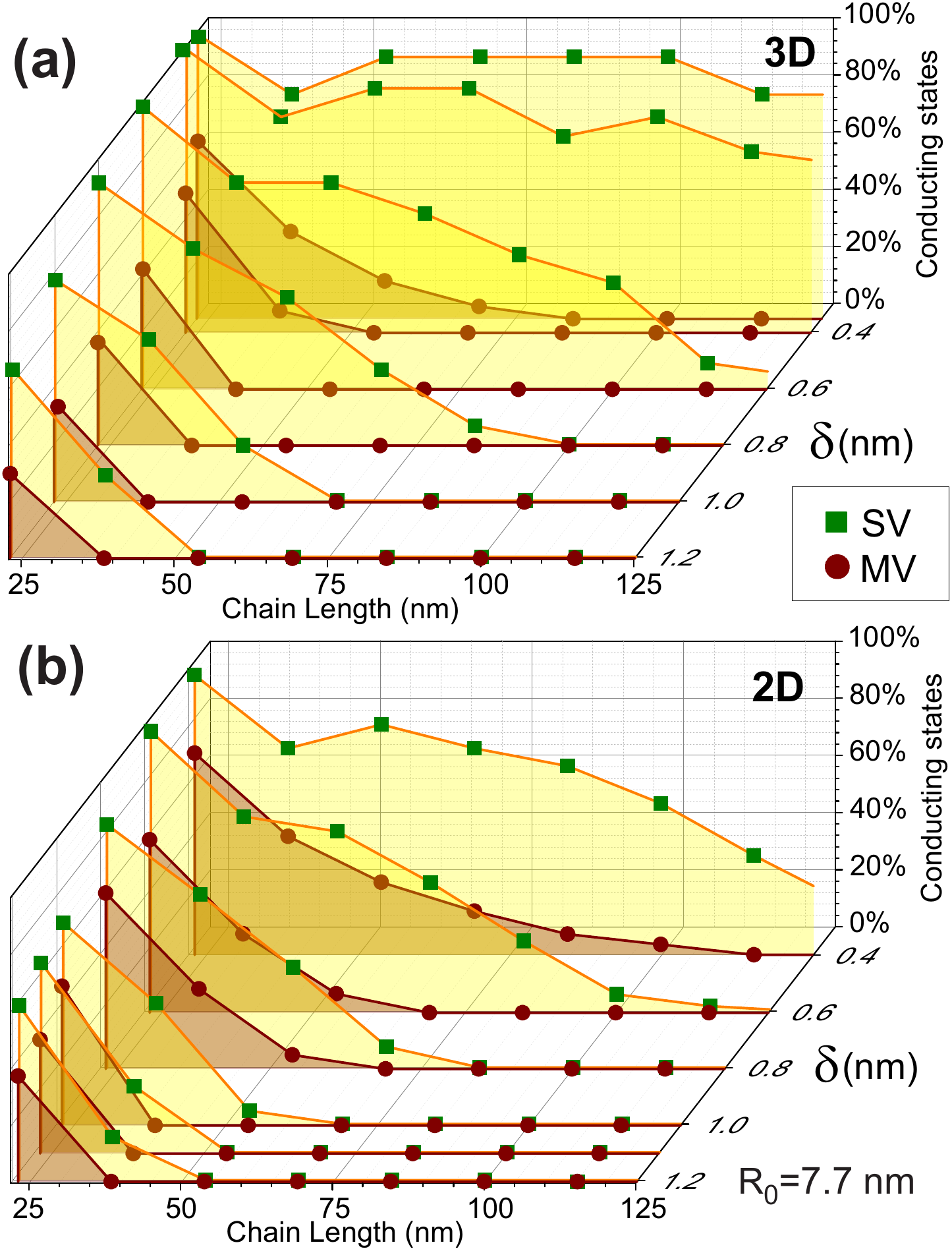}
		\caption{\label{fig:fraction}(Color online)
Fraction of conducting states as a function of donor chain length ($L$ and the disorder radius ($\delta$) for the single valley (SV) and multi-valley (MV) models. The 3D disorder model (a) and the 2D disorder model (b) were considered. Results are presented for a distance of $7.7$ nm between target implantation points.
  }
		\end{figure}

\section{General trends for 2D and 3D Models \label{sec:Trends}}

We may observe in Fig.~\ref{fig:Densities} that $\Lambda$ oscillates as a function of $R_0$, but an overall increase is noted. These oscillations are not due to statistical uncertainties -- all data is well converged beyond 1\% of the average. The increase in conductivity with a larger interdonor distance is
counter-intuitive and deserves further attention.

We point out two effects that impact the transport performance of donor chains under changes in donor density (or $R_0$). ​First, chains with ​strong enough​ hopping ​coupling​ connecting all​ pairs of ​​consecutive ​donors  ​clearly ​constitute​ ​​good​ electronic transport channels, ​thus ​larger values of $\Lambda$. This​ feature is favoured by ​smaller ​target ​interdonor distances ​in the​ fabrication of conductive chains. On the other hand, assuming first nearest neigh​​bor hoppings, ​a single​ interdonor​ pair​ distance​ ​leading to ​a ​strongly destructive interference​ in the coupling lowers the whole chain's ​conductivity​, ​​limited by this link​ alone​. ​
In terms of​​ holding transport​ along the same physical length L, a larger number of donors ​(smaller $R_0$​) probabilistically ​favours the​ formation of ​one or more ​weak link​s $(t\ll t_0)$​ and interrupting the transport flow. Th​ese​ effect​s​ ​contribute​ in ​​opposite direction​s​, ​to the conductivity​ changes with density: It is​ expected  that the competition between the​se ​two ​opposing ​effects induces some ​​non-monotonicity,
and leads to fluctuations due to the hardly predictable interference pattern ​of the tunnelling ​among donors​.

Figure~\ref{fig:Densities} also highlights that the 2D disorder model leads to longer conducting chains, compared to 3D disorder.
But the difference between these models is not too large, revealing that a three dimensional diffusion is not significantly more damaging to transport than the 2D imprecision.

In the multi-valley model at fixed $R_0= 7.7$~nm, the trend   $\Lambda_{2D} > \Lambda_{3D}$ is preserved  for all considered disorder parameters $\delta>0.4$nm, as illustrated in Fig. \ref{fig:zeta} (b) and (d).  In order to appreciate  valleys interference effects, results within the single valley model are also presented in Fig. \ref{fig:zeta}, showing that the general decay of $\Lambda$ with disorder is similar to the multi-valley description. Note that, for each $\delta$, single valley models lead to larger values of $\Lambda$ as compared to multi-valley,  consistent with  destructive valley interference affecting the hopping.
 We remark that, for this particular $R_0$, the 3D disorder model sustains larger $\Lambda$ values [See Fig.~\ref{fig:zeta} (a) and (c)]. It is possible that this ordering inverts at some  value of $R_0$ as in the multi-valley case, a question that is less relevant since the single valley model is not realistic.

Figure \ref{fig:AllDos-77A} presents the density of states (DOS) averaged over an ensemble of $10^4$ samples of chains with $\approx 500$ nm, differentiating single valley and multi-valley models. No significant differences are observed comparing the two models of disorder (2D and 3D). On the other hand, effects coming from valley interference may be observed in the DOS, where we note that the single valley DOS for small $\delta$ shows close similarities   to that of 1D ordered chains, while, as $\delta$ increases, the energy range of the spectrum widens and the peak related to the 1D van Hove-like peak lowers and spreads.  The multi valley DOS has no remanent features of the ordered 1D character, no memory of van Hove singularities at  the edges of the energy spectrum.  It is possible to observe that the multi valley DOS presents sharp oscillations for small $\delta$ that become smoother as $\delta$ increases.

Note that the DOS in the multi valley cases increases towards the band center as compared to the respective single valley DOS, which increases toward  the edges. In all cases the DOS has a peak at the center of the spectrum, i.e. the isolated-donor energy $\epsilon=0$. It is long known that in the case of purely off diagonal disorder in 1D the state at the center of the band is not exponentially localized.\cite{Soukoulis1981} While it still decays with distance, it is slower than the decay in Eq. \ref{eq:loc}. This favors the existence of longer localization lengths around the center of the spectrum.

While the  mobility edge is reduced in multi-valley materials due to valley interference, this accumulation of states near the band center favors transport. In our simulations we did not find any instances in which the latter effect is more relevant than the first. In other words, the number of conducting states is always larger in single valley materials.

 A comprehensive summary of our results is given in  Figure \ref{fig:fraction}. We analyze localization properties of chains with fixed target donors distance $\bf{R}_0$ along [110]. Ensembles are grouped by disorder $\delta$ and length $L$, where $\delta$ characterizes 2D or 3D position distributions and consequently tunnel coupling distributions. Single valley and multi-valley models are also treated independently. Overall we studied about 200 ensembles, corresponding to a fixed combination  $\{\delta, L,$ dimension of disorder (2D or 3D), valley multiplicity (single or multi)$\}$. The conduction character of each statistical realization is verified according to our criterion. The data points in the figure give the fraction of conducting samples in the respective ensemble.

\section{Discussion and conclusions}

Our study reveals non-trivial and sometimes unexpected consequences of fabrication parameters (${\bf R_0}$ and $L$) and fabrication control ($\delta$) on the transport behavior of 1D donor chains in silicon. For example, the non-monotonicity of the limiting length sustaining conduction $(\Lambda)$ is likely to persist even for more realistic models.
Our results are consistent with relatively long conducting chains reported in experiments,~\cite{Weber2012} even in disordered cases driven by valley interference effects (Fig. \ref{fig:Densities}). Distances between donors are reasonably controllable within STM-tip deposition techniques while the disorder radius is continually reduced with the development of these techniques.~\cite{Eng2006,Hu2012,Hwang2013,Shamim2014,Hsueh2014}

In our model, the Coulomb electron-electron correlations is not explicitly included. However, in the context of low dimensional systems like P-donors arrays in Si, explicit inclusion of electron-electron correlations are known to affect the electronic behavior and may eventually dominate the transport behavior.~\cite{Shamim2014} Nonetheless, the trends found here are expected to contribute to highly correlated chains, for which a detailed inclusion of the geometrical disorder and multi-valley effects may not be trivial.

\section{Acknowledgements}
We are indebted to Sergio Queiroz  for the many discussions on the TMA.
This work is part of the Brazilian National Institute for Science and Technology on Quantum Information.
The authors also acknowledge partial support from FAPERJ, CNPq and CAPES.
\vfill

\bibliography{Bibliography}

\bigskip\bigskip
\appendix
\section{\label{app:LyapExp} Computing Lyapunov Exponents and Integrated Density of states (IDOS) - 1D chain with off-diagonal disorder}

Considering the linear mapping presented in Equation \ref{eq:Mapping}, numerical implementation faces two main difficulties. In obtaining a final Lyapunov vector (LV) $\widehat{\Phi}_{L-1}$, one have to handle the product $\widehat{P}_L \cdot \widehat{\Phi}_{0}$ that diverges exponentially dominated by the Lyapunov maximum exponent (LME)--exceeding the overflow limit. Other difficult is related with the fact that the $\widehat{\Phi}_{L-1}$ asymptotic direction is given by the Oseledec subspace related with the LME, i.e., for any initial vector $\widehat{\Phi}_{0}$, the angle between final Lyapunov vectors will tends to zero leading to determinant precision problems.~\cite{Crisanti1993}

In this work, the solution for this difficulties lies in extract the increase and decrease of LV associated with the LME and the LmE (Lyapunov minimal exponent), respectively. For this purpose we made use of the Gram-Schmidt orthonomalization procedure (GSOP)--in the usual diagonal disorder context this procedure is implemented after $m$ steps but since in our problem the LVs diverge faster it will be done after each step. Although the procedure will be more expensive computationally it will impact in a very convenient way to determine the IDOS by the node counting technique.

We begin the GSOP initializing the LV, i.e., we randomly choose a initial set of orthonormal LVs: $\left(\widehat{\Phi}_{0}^{(1)},\widehat{\Phi}_{0}^{(2)}\right)$. After each step we extract the modulus of these LV and this set are orthonormalized again, for a $k+1$ step,
\begin{subequations}
\label{eq:GSOP}
\begin{eqnarray}
\widehat{\Phi}_{k+1}^{(1)}=\frac{\widehat{T}_{k+1} \cdot \widehat{\Phi}_{k}^{(1)}}{M^{(1)}_{k+1}}, \\
\widehat{\Phi}_{k+1}^{(0)}=\frac{\left[\widehat{1}-\widehat{\Phi}_{k+1}^{(1)}\widehat{\Phi}_{k+1}^{(1)\dagger}\right]\widehat{T}_{k+1} \cdot \widehat{\Phi}_{k}^{(0)}}{M^{(0)}_{k+1}},
\end{eqnarray}
\end{subequations}
here ($M^{(p=1,0)}_{k+1}$) is the modulus of each LV before the normalization procedure. The LME ($p=1$) and LmE ($p=0$) will be given by,
\begin{equation}
\gamma^{(p)}=\frac{1}{L}\sum^{L-1}_{i=0}\frac{\ln\left[M^{(p)}_{i}\right] }{\left|{\bf R}_i\right|}
\end{equation}
where $L$ is the chain length and ${\bf R}_i$ is the vector connecting a pair of nearest neighbors donors ($i$ and $i+1$). Although we apply a orthonormalization procedure to the set of LVs it is good to clarify that this set are not necessarily orthogonal, the goal here is decrease the influence of the LME in determining LVs associated with other Lyapunov exponents.

In our model, the $\textrm{n}^{\textrm{th}}$ eigenstate must have n nodes and given this, the nunber of states below the energy given by $\textrm{n}^{\textrm{th}}$ eigenstate will be precisely n.  By the node counting technique we determine the IDOS by the ratio of the wavefunction amplitudes ($\phi_n$) presented in Equation \ref{eq:TMequationRel}, i.e.,
\begin{equation}
IDOS(E)=\frac{1}{L}\sum_{i=0}^{N-1}\Theta\left[-\frac{\phi_i(E)}{\phi_{i+1}(E)}\right],
\end{equation}
$\Theta$ is the Heaviside theta. By the IDOS the density of states is easily obtained.

\end{document}